\documentclass[twocolumn,twosided,english,nofootinbib,aps]{revtex4-1}
\usepackage{color}
\usepackage{amsmath}
\usepackage{graphicx}
\usepackage{amssymb}
\usepackage{esint}
\usepackage[
hyperfootnotes=true,
unicode=true, 
 bookmarks=true,bookmarksnumbered=false,bookmarksopen=false,
 breaklinks=false,pdfborder={0 0 1},backref=false,colorlinks=true]
 {hyperref}

\makeatletter

\@ifundefined{textcolor}{}
{%
 \definecolor{BLACK}{gray}{0}
 \definecolor{WHITE}{gray}{1}
 \definecolor{RED}{rgb}{1,0,0}
 \definecolor{GREEN}{rgb}{0,1,0}
 \definecolor{BLUE}{rgb}{0,0,1}
 \definecolor{CYAN}{cmyk}{1,0,0,0}
 \definecolor{MAGENTA}{cmyk}{0,1,0,0}
 \definecolor{YELLOW}{cmyk}{0,0,1,0}
 }

\usepackage{hyperref}
\hypersetup{
    colorlinks,%
    citecolor=blue,%
    filecolor=blue,%
    linkcolor=blue,%
    urlcolor=blue
}

\makeatother

\usepackage{enumitem}

\newcommand{\be}{\begin{eqnarray}}
\newcommand{\ee}{\end{eqnarray}}

\begin{document}


\title{
Runaway potentials and a massive goldstino 
}

\author{Fotis Farakos}
\email{fotios.farakos@kuleuven.be}
\affiliation{ KU Leuven, Institute for Theoretical Physics, \\
			Celestijnenlaan 200D, B-3001 Leuven, Belgium}


\begin{abstract}

We study N=1 globally supersymmetric theories on 
runaway backgrounds arising 
from scalar potentials 
with a slope characterized by a scale $M = |V/V'|$. 
We find that, 
under mild assumptions, 
there always exists a massive goldstino in the low energy effective theory. 
In the simplest models the effective mass of such a fermion is of order $\sqrt{V} / M$, 
or of order $V / M^2 \mu$ when a seesaw mechanism takes place, 
where $\mu$ is a scale that characterizes the masses of other heavy fermions.

\end{abstract}


\maketitle

\section{Introduction}

Runaway potentials have received a renewed interest 
because of the controversy surrounding (meta) stable de Sitter vacua, 
not only in String Theory \cite{Danielsson:2018ztv,Kallosh:2018nrk}, 
but also within Quantum Gravity in general \cite{Dvali:2018fqu,Dvali:2018jhn}. 
Moreover, 
a {\it de Sitter conjecture} was put forward in \cite{Obied:2018sgi}, 
restricting the form of the scalar potential $V$ in low energy effective theories, 
and excluding de Sitter critical points. 
For generic runaway potentials, for a {\it positive} $V$, one has 
\be
\label{SC1}
\frac{|\nabla_{\phi} V|}{V}  = \frac{1}{M(\phi)} > 0 \, . 
\ee
In equation \eqref{SC1}, 
$M(\phi)$ is some field-dependent mass scale 
and $\nabla_{\phi} V$ refers to the derivative with respect to scalars with canonically normalized kinetic terms. 
In particular, the authors of \cite{Obied:2018sgi} have put a bound on $M(\phi)$ and have correlated it explicitly with $M_P$. 
However, 
if one simply studies runaway potentials, such restriction is unwarranted. 
Indeed, it was subsequently pointed out that the bound on $M$ set by \cite{Obied:2018sgi} may be 
difficult to reconcile with the Standard Model Higgs potential \cite{Denef:2018etk}, 
and further {\it refinements} of the de Sitter conjecture have been proposed \cite{Garg:2018reu,Ooguri:2018wrx,Andriot:2018ept,Banlaki:2018ayh,Andriot:2018mav} 
(for recent reviews see \cite{Andriot:2019wrs,Palti:2019pca}). 
The implications of such a conjecture on inflation 
have been summarized in \cite{Kehagias:2018uem,Agrawal:2018own}, 
and the study of the implications on late-time cosmology \cite{Agrawal:2018own,Chiang:2018jdg,Cicoli:2018kdo,Olguin-Tejo:2018pfq,Agrawal:2018rcg} supports 
a quintessence-type behavior \cite{Wetterich:1987fm,Ratra:1987rm,Caldwell:1997ii} for the low energy 
supergravity theory \cite{Hellerman:2001yi,Brax:1999gp,Copeland:2000vh,Akrami:2017cir}. 
For further recent developments on the de Sitter KKLT construction \cite{Kachru:2003aw} see 
\cite{Akrami:2018ylq,Bena:2018fqc,Gautason:2018gln,Hamada:2018qef,Kallosh:2019axr,Kallosh:2019oxv,Gautason:2019jwq,Hamada:2019ack,Carta:2019rhx}, 
and for discussions on other possible counter-examples to 
the de Sitter conjectures 
see \cite{Weissenbacher:2019bfb,Cordova:2018dbb,Cribiori:2019clo,Blaback:2019zig}.

Our approach in this article is to remain agnostic about the validity of the de Sitter conjectures, 
and study instead directly the implications of such 
a runaway  background on the low energy N=1 supersymmetric theory.  
In other words, 
we are solely interested in the runaway behavior implied by \eqref{SC1} 
and its implications. 
In particular, as $M_P$ does not explicitly appear in \eqref{SC1}, this condition may hold also when gravity is decoupled, 
assuming that $M$ is not explicitly correlated to $M_P$. 
Let us therefore assume that the condition \eqref{SC1} holds for a globally supersymmetric field theory. 
Then, 
at least one scalar (say $\phi$) of some supermultiplet will have a runaway behavior, 
implying that the scalar potential part of the Lagrangian has the form 
\be
\label{LIN}
{\cal L}_V = -V = -V_0 - c \, \phi + \dots 
\ee 
Here we assume for simplicity that $\phi$ and $c$ are real without loss of generality. 
Clearly, 
if this background is to satisfy \eqref{SC1} in a non-trivial way (e.g. $V \ne 0$), 
then the  scalar background will develop a runaway time-dependent behavior 
and supersymmetry will be generically broken \cite{Giudice:1999am,FVP}. 
However, 
assuming the breaking is spontaneous, 
then such theory will still be invariant under the supersymmetry transformations. 
For the scalar $\phi$, 
the supersymmetry transformations will generically have the form 
\be
\phi  \to \phi + \epsilon \psi  + \dots 
\ee 
where the fermion $\psi$ is the superpartner of $\phi$ and $\epsilon$ is the supersymmetry parameter. 
Then the scalar potential part of the Lagrangian,
that is \eqref{LIN}, 
will produce the following term under a supersymmetry variation 
\be
\label{dV} 
 - c \, \epsilon \psi - c \, \overline \epsilon \overline \psi  \, . 
\ee
The only term that can cancel \eqref{dV} is a term of the form 
\be
\label{LGpsi} 
{\cal L}_{G-\psi} = - c \, G^\alpha \psi_\alpha - c \, \overline G_{\dot \alpha} \overline \psi^{\dot \alpha} \, , 
\ee
where $G$ is the goldstino, 
that is, a linear combination of the physical fermions of the low energy theory, 
that transforms under supersymmetry as 
\be
\label{deltaG} 
G_\alpha \to G_\alpha + \epsilon_\alpha + \dots  
\ee
The linear combination of the physical fermions that contribute to the goldstino 
is uniquely fixed by requiring that 
the latter is the only fermion with a non-trivial shift 
in its supersymmetry transformation, 
when it is evaluated on the background \cite{Giudice:1999am}. 
In \eqref{LGpsi} and in \eqref{deltaG} for simplicity we have set the supersymmetry breaking scale to unity, 
but we will restore it when we study supersymmetric chiral models in the next sections.

Let us now inspect the meaning of a term of the form \eqref{LGpsi} in more detail. 
Generically, 
either the goldstino will be identified with $\psi$, 
and therefore have a {\it Majorana mass}, 
or, if it is not completely aligned with $\psi$, 
then the two fermions will share a {\it Dirac mass}. 
In both cases however, 
independent of the details, 
a term of the form \eqref{LGpsi} will signal that the goldstino has acquired an effective mass. 
As  we will see in the rest of this article, 
the existence of a massive goldstino in runaway dynamical backgrounds 
is a generic feature of globally supersymmetric theories. 
Our findings can be contrasted to vacua with stabilized scalars (or with flat directions), 
which violate \eqref{SC1}, 
and would give rise to a {\it massless} goldstino. 
Therefore supersymmetric theories on runaway backgrounds lead to distinct phenomenological implications that deserve a dedicated study.

Before closing this section, 
let us note that other types of {\it massive goldstini} 
have been studied in supersymmetry within different contexts 
in \cite{Argurio:2011hs,Cheung:2010mc,Benakli:2007zza,Benakli:2014daa,Liu:2016idz,Bardeen:1981df}. 
These massive fermions have different physical origin from the massive goldstino we study here. 
For example, 
the massive fermions studied in \cite{Cheung:2010mc,Argurio:2011hs} are {\it not} true goldstini (they do not shift), 
rather they are the orthogonal fermions to the goldstino that generically become massive, 
whereas the goldstino itself remains massless. 
The breaking of supersymmetry in \cite{Cheung:2010mc,Argurio:2011hs} is of course spontaneous and it originates from multiple sectors. 
Moreover, 
the pseudo-goldstini studied in \cite{Liu:2016idz} 
are related to 
the explicit breaking of supersymmetry.

This article is organized as follows: 
In the second section we discuss generic time-dependent scalar backgrounds 
and show how Yukawa couplings give rise to effective fermionic masses. 
In section 3 we elaborate on chiral models with a single chiral superfield, 
and we exemplify our general discussion from the introduction 
with two explicit models that contain effective goldstino mass terms. 
In the fourth section we discuss chiral models with multiple chiral superfields and study 
the fermion mass matrix on time-dependent scalar backgrounds. 
We close in section 5 with a short discussion of our findings and future directions.

\section{Effective field theory on time-dependent scalar backgrounds}

In supersymmetric field theories the fermions generically appear with Yukawa-type couplings to the scalars \cite{Wess:1992cp,FVP}. 
As a result, 
non-trivial time-dependent scalar backgrounds will generically give rise to effective fermion masses, 
even for the goldstino \cite{Giudice:1999am}. 
Therefore, 
before we introduce supersymmetric theories we would like to set the stage for the 
study of these effective fermionic mass terms. 
To this end, 
consider a field theory with a fermion $\chi$ and a real scalar $\phi$ 
described by a Lagrangian\footnote{We use the conventions of \cite{Wess:1992cp}.}  
\be
\label{LEX}
\begin{aligned}
{\cal L} = & - \frac12 \partial_m \phi \partial^m \phi 
- i \overline \chi \overline \sigma^m \partial_m \chi 
\\
& - \frac12 m(\phi) \chi^2 - \frac12 m(\phi) \overline\chi^2  - V(\phi) \, . 
\end{aligned}
\ee
Let us split the scalar $\phi$ as 
\be
\phi = \phi_\mathbb{B} + \delta \phi \, ,  
\ee
where $\phi_\mathbb{B}$ will serve as the background for the scalar while $\delta \phi$ describes the fluctuations around such background. 
By varying the Lagrangian 
we find that the equations satisfied by the scalar background have the form 
\be
\label{split-phi}
\partial^m \partial_m \phi_\mathbb{B} = V'(\phi_\mathbb{B}) \, . 
\ee
Clearly, 
if we wish to study the theory around a critical point then the background will satisfy $V'(\phi_\mathbb{B}) = 0$. 
However, 
if we are not explicitly interested in the critical points of the scalar potential, 
we can still 
replace the split form of the scalar \eqref{split-phi} into the Lagrangian \eqref{LEX} 
and study the effective theory of the fluctuations around the background $\phi_\mathbb{B}$. 
Then, 
the total action describing the system will have the form 
\be
\label{SPSP}
S = S|_{\mathbb{B}}  +  S_\text{eff} \, . 
\ee
The background part is given by 
\be
S|_{\mathbb{B}}  = \int d^4 x \left( - \frac12 \partial_m \phi_\mathbb{B} \partial^m \phi_\mathbb{B}  - V(\phi_\mathbb{B}) \right) \, . 
\ee 
The effective part of the action $S_\text{eff} = \int d^4 x \, {\cal L}_\text{eff}$, 
containing the fluctuations, 
is given in terms of the Lagrangian 
\be
\label{LEX'}
\begin{aligned}
{\cal L}_\text{eff} = & - \frac12 \partial_m \delta \phi \, \partial^m \delta \phi 
- i \overline \chi \overline \sigma^m \partial_m \chi 
\\
& - \frac12 m_\text{eff} \, \chi^2 - \frac12 m_\text{eff} \, \overline\chi^2  - V_\text{eff} \, , 
\end{aligned}
\ee
where 
\be
V_\text{eff} = \sum_{n=2}^{\infty} \frac{1}{n !} V^{(n)}\Big{|}_{\phi_\mathbb{B}} \delta \phi^n \, , 
\ee
and 
\be
m_\text{eff} = m(\phi_\mathbb{B}) +  \sum_{n=1}^{\infty} \frac{1}{n !} m^{(n)}|_{\phi_\mathbb{B}} \delta \phi^n \, . 
\ee
Here we used the notation $V^{(n)} = \partial^n V / \partial \phi^n$ and similarly for the mass $m$, 
and with $``|_\mathbb{B}"$ we mean that the quantity is evaluated on the background. 
Notice that there are no linear terms with respect to $\delta \phi$ in the effective Lagrangian except of the Yukava couplings. 
This is happening because the background satisfies the bosonic equations of motion. 
Therefore, because of the splitting \eqref{SPSP}, 
the effective action  will contain the Gaussian terms and higher order interactions, 
and  can be treated within perturbation theory.

Let us turn to the fermionic mass matrix on such a background. 
The fermionic equations of motion can be now derived from the effective action $S_\text{eff}$. 
We vary $S_\text{eff}$ with respect to $\overline \chi^{\dot \alpha}$ and find 
\be
\label{EQeff}
\!\!\!\!\!\! - i \overline \sigma^m_{\dot \alpha \beta} \partial_m \chi^{\beta} 
=  \overline \chi_{\dot \alpha} \Big{[} m(\phi_\mathbb{B}) +  \sum_{n=1}^{\infty} \frac{1}{n !} m^{(n)}|_{\phi_\mathbb{B}} \delta \phi^n \Big{]}  .   
\ee
Equation \eqref{EQeff} describes the propagation of a massive spinor with mass given by evaluating the $m(\phi)$ on the background, 
whereas the rest of the terms that appear describe interactions of the Yukawa type with the scalar fluctuations.

We conclude that the effective theory that arises from \eqref{LEX} 
when it is studied around a scalar field background $\phi_\mathbb{B}$ that satisfies the bosonic equations of motion 
describes a massive spinor with effective mass given by 
\be
m(\phi_\mathbb{B}) \, . 
\ee
Clearly our discussion here can also be extended to theories with multiple fermions 
and we will study such supersymmetric theories in the fourth section.

\section{Single chiral superfield}

Let us first discuss a supersymmetric field theory with 
a single chiral superfield 
$\Phi = A + \sqrt 2 \theta \chi + \theta^2 F$, 
where $A$ is a complex scalar, 
$\chi$ is the fermion superpartner of $A$ 
and $F$ is the complex scalar auxiliary field. 
We study the Lagrangian 
\be
\label{CH1}
{\cal L} = \int d^4 \theta \, \Phi \overline \Phi + \left\{ \int d^2 \theta \, W(\Phi) + c.c. \right\} \, .  
\ee
Here $W(\Phi)$ is the superpotential, 
which is a holomorphic function of $\Phi$. 
Once we reduce \eqref{CH1} to components and integrate out the auxiliary field we find 
\be
\label{CH1comp}
\begin{aligned}
{\cal L} = & -\partial_m A \, \partial^m \overline A 
- i \overline \chi \overline \sigma^m \partial_m \chi  
\\ 
& - \frac12 ( W'' \chi^2 + \overline W'' \overline \chi^2 )
- V \, , 
\end{aligned}
\ee
where the scalar potential has the form 
\be
V = W' \overline W' \, , 
\ee 
with $W' = \partial W / \partial A$.

We wish to study the effective theory of the fluctuations 
around a non-trivial time-dependent background $A_\mathbb{B}(t)$ 
that satisfies the scalar equations of motion. 
To this end we can split $A$ as 
\be
A = A_\mathbb{B}(t) + \delta A \, , 
\ee
where $\partial^2 A_\mathbb{B} = V'(A_\mathbb{B})$. 
Assuming that the scalar potential satisfies the condition \eqref{SC1} gives generically 
\be
\label{1overM}
 \frac{W''(A_\mathbb{B})}{W'(A_\mathbb{B})} = {\cal C}(A_\mathbb{B})  \, ,  
\ee 
with ${\cal C}(A_\mathbb{B})$ related to the scale $M$ as $|{\cal C}(A_\mathbb{B})| = M^{-1}$. 
As a result, 
for the scalar field background $A_\mathbb{B}$ we should assume 
\be
\label{Wne0}
W'(A_\mathbb{B}) \ne 0 \, , \quad W''(A_\mathbb{B}) \ne 0 \, . 
\ee  
Clearly such background will break supersymmetry 
as the energy density of the system evaluated on the background 
is non-vanishing and given by \cite{Giudice:1999am} 
\be
\rho\Big{|}_\mathbb{B} = \Big{|}\frac{\partial A_\mathbb{B}}{\partial t} \Big{|}^2 + |W'(A_\mathbb{B})|^2 > 0 \, . 
\ee 
The supersymmetry transformation of the fermion $\chi$ 
evaluated on such background 
is a shift 
\be
\!\!\!\! \delta_{\rm SUSY} \chi_\alpha \Big{|}_\mathbb{B}  = - \sqrt 2 \, \overline W'(A_\mathbb{B}) \epsilon_\alpha  
+ i \sqrt{2} \sigma^{0}_{\beta \dot{\beta}} \overline{\epsilon}^{\dot{\beta}} \frac{\partial A_\mathbb{B}}{\partial t}  \, , 
\ee
and it describes the goldstino.

As we have explained in the previous section, 
to find the effective mass of the fermion we only have to evaluate its mass term given in \eqref{CH1comp} on the bosonic background. 
We find 
\be
m_{\chi}(A_\mathbb{B}) = W''(A_\mathbb{B}) \ne 0 \, , 
\ee
therefore the goldstino is {\em massive}. 
Evaluating \eqref{1overM} on the background 
gives 
\be
\label{SingleM} 
|m_{\chi}|^2 = \Big{|} \frac{W'' W'}{W'} \Big{|}^2 = V |{\cal C}|^2 \, . 
\ee 
As we see the mass of the goldstino is related only to the 
value of the scalar potential evaluated on the background 
and to the scale $M$ (via ${\cal C}$) that characterizes the slope. 
Identifying $|{\cal C}(A_\mathbb{B})| = M^{-1}$ we find 
\be
\label{mVM} 
|m_{\chi}| = \sqrt{V} / M \, . 
\ee
The relation \eqref{mVM} will appear also in models that include more than one fermions as we will see later.

Let us now stress that 
if we take the limit 
\be
W'(A_\mathbb{B}) = \text{finite} \, , \quad W''(A_\mathbb{B}) \to 0 \, , 
\ee
which is equivalent to 
\be
V(A_\mathbb{B}) = \text{finite} \, , \quad {\cal C}(A_\mathbb{B}) \to 0 \, , 
\ee
we find 
\be
m_{\chi}(A_\mathbb{B}) \to 0 \, . 
\ee
This happens because in such background 
the scalar is stabilized while supersymmetry is broken, 
therefore 
the goldstino essentially becomes massless. 
In the example we studied here as there is only one fermionic mode in the effective theory this observation is trivial. 
However, 
when we study models with multiple fermions, 
we will identify the eigenvalue of the fermionic mass matrix corresponding to the goldstino 
as the one that vanishes in the limit ${\cal C}(A_\mathbb{B}) \to 0$. 
Such limit might not be physical in a generic setup, 
however we will only employ it as a {\it formal} limit 
that will help us identify the goldstino effective mass on a non-trivial background.

The goldstino was also studied in a non-trivial background in 
\cite{Kahn:2015mla}\footnote{Supergravity models with a time dependent background have been also studied in \cite{Koehn:2012te} within a different context.}, 
directly within a non-linear realization of supersymmetry, 
where it was assumed that terms including $\dot \phi$ acquire non-vanishing {\it vevs}. 
The goldstino was found to be massless in such setup. 
Nevertheless, 
terms contributing to a goldstino effective mass may appear in \cite{Kahn:2015mla} with order $\ddot \phi$. 
The latter terms were ignored in \cite{Kahn:2015mla} 
as being highly suppressed, 
therefore a detailed study of these terms is required to compare with our analysis here. 
Our findings however are in complete agreement with \cite{Giudice:1999am}, 
where it is explained how the goldstino acquires an effective mass during reheating.

Before we turn to generalizations with multiple fermions, 
let us discuss two simple examples with a single chiral superfield.

\subsection{Displacement from supersymmetric vacuum}

As a simple example we can discuss the superpotential 
\be
W = \frac12 m \Phi^2  \, . 
\ee
This model has a massive fermion with mass 
\be
m_\chi = m \, , 
\ee
which is {\it independent} of the background. 
Clearly on the vacuum where $\langle A \rangle=0$ supersymmetry is unbroken, 
thus the fermion is not the golstino and it is massive,  
forming a massive multiplet with the scalar. 
However, 
we can assume that the scalar is displaced from its true vacuum to some position $A^*$ at time $t^*$, 
and therefore an effective theory can be constructed for the non-trivial background of the form 
\be
\ddot A_\mathbb{B} = - m^2 A_\mathbb{B}   \, , \quad A_\mathbb{B}(t=t^*) = A^* \ne 0 \, . 
\ee
We then find 
\be
V|_\mathbb{B} = m^2 |A_\mathbb{B}|^2 \, , \quad {\cal C}|_\mathbb{B} = \frac{1}{A_\mathbb{B}} \, . 
\ee
In such case we have the general relation \eqref{SingleM} satisfied trivially 
\be
\Big{|} m_\chi (A_\mathbb{B}) \Big{|}^2 = m^2 =  m^2 |A_\mathbb{B}|^2 \frac{1}{|A_\mathbb{B}|^2} \, ,  
\ee
giving a background with broken supersymmetry and a massive goldstino.

\subsection{Runaway potential}

The second example we would like to study is a 
supersymmetric theory with a runaway potential 
that is given by 
\be
W = W_0 \, \text{e}^{- \Phi / m} \, , 
\ee
where $m$ is a real positive constant. 
In this setup the scalar potential has a runaway behavior 
\be
V = \frac{|W_0|^2}{m^2} \text{e}^{- (A + \overline A) / m} \, , 
\ee 
and supersymmetry is in a spontaneously broken phase. 
For the slope parameter we find 
\be
{\cal C} = - \frac1m\,. 
\ee
Assuming that we study the theory around a non-trivial background $A_\mathbb{B}$, 
we find 
\be
|m_\chi(A_\mathbb{B})|^2 
= \frac{|W_0|^2}{m^4} \text{e}^{- (A_\mathbb{B} + \overline A_\mathbb{B}) / m} 
= V(A_\mathbb{B}) |{\cal C}|^2 \, , 
\ee
in agreement with our general discussion.

\section{Multiple chiral superfields}

We can now extend our discussion to a theory with multiple 
chiral multiplets that are described by the superfields $\Phi^I$ with 
$\theta$-expansion 
\be
\Phi^I = A^I + \sqrt 2 \theta \chi^I + \theta^2 F^I \, . 
\ee 
The $A^I$ describe the physical complex scalars, 
the $\chi^I$ are Weyl spinors that describe the fermions, 
and the $F^I$ are the auxiliary fields. 
The supersymmetry transformations of the fermions are 
\be
\label{SUSYGEN}
\delta \chi^I_\alpha = \sqrt 2 F^I \epsilon_\alpha  
+ i \sqrt{2} \sigma^{a}_{\beta \dot{\beta}} \overline{\epsilon}^{\dot{\beta}} \partial_a A^I  \, . 
\ee
When we study the theory around a background where $F^I \ne 0$ 
or $\dot A^I \ne 0$ 
these fermions contribute collectively to the goldstino.

The most general chiral model (up to two derivatives) 
has the form \cite{Wess:1992cp} 
\be
\label{SCM}
{\cal L} = \int d^4 \theta \, K\left( \Phi,\overline \Phi \right) + \left\{ \int d^2 \theta \, W(\Phi) + c.c. \right\} \, . 
\ee
The holomorphic function $W(\Phi)$ is the superpotential, 
and the function $K$ is real and it is the K\"ahler potential. 
The K\"ahler metric is defined as 
\be
g_{IJ} = K_{I \overline J} = \frac{\partial^2 K}{\partial A^I \partial \overline A^{\overline J}} \, . 
\ee
The connection is given by $\Gamma^K_{IJ} = g^{\overline L K} \partial_{I} g_{J \overline L}$. 
Clearly one can choose a specific set of $A^I$ 
such that at a given point $A^I_{*}$ in field space 
it will hold $\Gamma^K_{IJ}|_{*} = 0$, 
however the derivatives of the connection evaluated at that point will not generically vanish. 
Note that the derivatives of the connection $\Gamma^K_{IJ}$ 
contribute only to the four-fermion interactions in the component form of the Lagrangian \eqref{SCM}. 
To proceed with our discussion we assume from now on that 
\be
\label{flat}
g_{IJ}  = \delta_{IJ} \, , 
\ee
such that the K\"ahler manifold is flat. 
As we have explained this is not a very strict requirement 
as one can generically go to a coordinate system 
where \eqref{flat} will hold for small field excursions and higher order terms will be highly supressed. 
However, 
for clarity, 
we will impose \eqref{flat} in general, 
which then leads to a K\"ahler potential of the form 
\be
\label{FLTK}
K = \delta_{IJ} \Phi^I \overline \Phi^{\overline J} \, . 
\ee 
The component form of the Lagrangian \eqref{SCM}, 
with K\"ahler potential \eqref{FLTK}, reads 
\be
\label{GENGEN}
\begin{aligned}
{\cal L} = & - \delta_{IJ} \partial A^I \partial \overline A^{\overline J} 
- i \delta_{IJ} \overline \chi^{\overline J} \overline \sigma^m \partial_m \chi^{I}   
\\
&- \frac12 W_{IJ}  \chi^I \chi^J 
- \frac12 \overline W_{\overline I \overline J} \overline \chi^{\overline I} \overline \chi^{\overline J} - V \, , 
\end{aligned}
\ee
where 
\be
V = \delta^{IJ} W_I \overline W_{\overline J} \, . 
\ee
In \eqref{GENGEN} the auxiliary fields have been integrated out and take the values 
\be
F^I = \delta^{IJ} \overline W_{\overline J} \, . 
\ee
Notice finally that the fermion kinetic terms in \eqref{GENGEN} are canonical.

Let us now study the effective theory around a non-trivial background for the complex scalars $A^I$. 
Without loss of generality we can assume that the scalar that gives rise to the non-trivial background is $A^o$, 
while all other scalars are constant (on the background). 
We first split the scalars in their background values and their fluctuations as follows 
\be
A^I = A^I_\mathbb{B}(t) + \delta A^I \, ,   
\ee
and we see that the equations that define the bosonic background have the form 
\be
\ddot A^I_\mathbb{B}(t) =  - \delta^{IJ}  \frac{\partial V}{\partial \overline A^J} \Big{|}_\mathbb{B} \, .  
\ee
Then we split the scalar manifold coordinates as $I=(o,A)$ and we have 
\be 
\label{NTB}
\ddot A^A_\mathbb{B}(t) = 0 \, , \quad 
\ddot A^o_\mathbb{B}(t) =  - \frac{\partial V}{\partial \overline A^o} \Big{|}_\mathbb{B} \ne 0 \, . 
\ee 
In accordance to our previous discussions we  set 
\be 
\frac{\partial V}{\partial A^o} \Big{|}_\mathbb{B} =  V_o \Big{|}_\mathbb{B} = c \ne 0 \, . 
\ee
The property of the background then leads to the condition 
\be 
\label{V_0} 
\delta^{IJ} W_{I o}  \overline W_{\overline J} \Big{|}_\mathbb{B} =  c \ne 0 \, , 
\ee
but because the other scalars $A^A$ are all stabilized we also have 
\be
\label{V_A}
V_A \Big{|}_\mathbb{B} = 0 \quad \to \quad  \delta^{IJ} W_{I A}  \overline W_{\overline J} \Big{|}_\mathbb{B} = 0 \, . 
\ee
We have used here the notation $V_A = \partial V / \partial A^A$. 
In addition, 
for \eqref{V_0} to hold we have 
\be
W_I \Big{|}_\mathbb{B} \ne 0 \, ,  \quad V \Big{|}_\mathbb{B} \ne 0 \, , 
\ee
and therefore supersymmetry is spontaneously broken.

We wish now to evaluate the mass of the goldstino $m_G$ on the non-trivial background \eqref{NTB}. 
One way to proceed would be to identify the goldstino as in \cite{Giudice:1999am}, 
then define the orthogonal fermions to the goldstino, 
and finally calculate the eigenvalue of the mass matrix corresponding to the goldstino on the non-trivial background. 
However, 
since we are interested {\it only} in identifying the effective mass of the goldstino there is a simpler 
method which we can follow. 
As we have explained, 
the fermionic mass matrix will have an eigenvalue $m_G$ that vanishes in the {\it formal} limit $c \to 0$. 
This limit of course has to be taken with care, 
because $m_G$ will not generically vanish if in such limit supersymmetry is allowed to be restored. 
We will therefore proceed by ascribing to the effective goldstino mass 
the mass matrix eigenvalue 
in the non-trivial background \eqref{NTB} that has the property 
\be
m_{G}\Big{|}_{c \to 0} \to 0 \, , \quad V \Big{|}_{c \to 0} \ne 0 \, . 
\ee 
With this strategy in mind, 
we will proceed to write down the fermion mass matrix. 
The fermionic masses read 
\be
\label{MASSterm}
- \frac12 \hat W_{oo}  \chi^o \chi^o  
- \hat W_{oA}  \chi^o \chi^A  
- \frac12 \hat W_{AB}  \chi^A \chi^B \, , 
\ee 
where the $\hat W_{IJ}$ are evaluated on the background \eqref{NTB}, 
that is 
\be
\hat W_{IJ}  = W_{IJ} \Big{|}_\mathbb{B} \, , \quad \hat W_{I}  = W_{I} \Big{|}_\mathbb{B} \, . 
\ee
There are now three possibilities depending on the type of supersymmetry breaking: 
\begin{enumerate}[label=\Alph*.] 

\item We have ${\hat W}_o = f \ne 0$ and ${\hat W}_A = 0$. 

\item We have ${\hat W}_o = 0$ and ${\hat W}_A = f_A \ne 0$. 

\item We have ${\hat W}_o =f_o \ne 0$ and ${\hat W}_A = f_A \ne 0$. 

\end{enumerate} 
From now on, 
without loss of generality we will assume that $f$, $f_A$ are real 
and that $c$ is real and positive. 
Notice that the $c$ is related to the scale $M$ appearing in \eqref{SC1} via 
\be
c = \frac{V_\mathbb{B}}{M} \, , 
\ee
and that 
\be
V_\mathbb{B} = \delta^{IJ} \hat W_I \overline{\hat W}_{\overline J} \, . 
\ee
We now turn to the study of the three possibilities A., B. and C.

\subsection{${\hat W}_o = f \ne 0$ and ${\hat W}_A = 0$}

We start with the possibility A. which also means that the 
fermion $\chi^o$ is completely aligned with the goldstino.  
The conditions \eqref{V_0} and \eqref{V_A} imply  
\be
{\hat W}_{oo} = \frac{c}{f} \ne 0 \, ,  \quad {\hat W}_{oA} = 0 \, . 
\ee 
Here the effective mass matrix is 
\be
\label{MAA}
m_{IJ} = 
\begin{pmatrix}
  \frac{c}{f} &  0   \\
    0 & {\hat W}_{AB}  
  \end{pmatrix} \, . 
\ee  
Clearly the goldstino is massive, 
with mass proportional to $c$, whereas the masses 
of the other fermions are independent 
and given by ${\hat W}_{AB}$. 
Notice that in particular for the goldstino mass we have 
\be
\label{MGOO}
m_G = \frac{c}{f} = \frac{V_\mathbb{B}}{M} \frac{1}{\sqrt{V_\mathbb{B}}} =  \frac{\sqrt{V_\mathbb{B}}}{M} \, , 
\ee
delivering the same result as the single chiral multiplet model.

\subsection{${\hat W}_o = 0$ and ${\hat W}_A = f_A \ne 0$}

We now turn to the possibility B. 
and rotate the superfields $\Phi^A$ such that only one of them 
has a non-vanishing value for the auxiliary field evaluated on the background. 
Namely we set 
\be
f_o=0 \, , \quad f_1 = f \ne 0 \, , \quad f_{a} = 0 \, , 
\ee 
where we further split the scalar manifold coordinates as $A = (1, a)$. 
The conditions \eqref{V_0} and \eqref{V_A} imply  
\be
{\hat W}_{oo} = {\rm m}  \, , \quad {\hat W}_{o1} = \frac{c}{f} \ne 0 \, ,  \quad {\hat W}_{11} = 0 \, , 
\ee
and 
\be
\quad {\hat W}_{1a} = 0 \, , \quad {\hat W}_{oa} = v_a  \, . 
\ee
We stress that the values of m and $v_a$ are unconstrained, 
we choose them however to be real and positive. 
Therefore we have the effective mass matrix for the fermions $(\chi^o, \chi^1,\chi^a)$ given by 
\be
\label{M1}
m_{IJ} = 
\begin{pmatrix}
  {\rm m} &  \frac{c}{f} & v_a  \\
   \frac{c}{f} & 0 & 0 \\ 
   v_b & 0 & {\hat W}_{ab}  
  \end{pmatrix} \, . 
\ee 
The equation that defines the eigenvalues $\lambda$ reads 
\be
\label{MLL}
\begin{aligned}
&\frac{c^2}{f^2} \det[\hat W_{ab} - \lambda \, \delta_{ab} ] 
\\
&+ \lambda \det\Big{[}\begin{pmatrix}
  {\rm m}-\lambda  & v_a  \\ 
   v_b & {\hat W}_{ab} - \lambda \, \delta_{ab}  
  \end{pmatrix}\Big{]} = 0 \, .  
\end{aligned}
\ee
Equation \eqref{MLL} is a higher order polynomial equation.  
Notice however that \eqref{MLL} does describe an eigenvalue that goes to zero as $c$ goes to zero. 
This is the eigenvalue that corresponds to the goldstino. 
To illustrate the behavior of this eigenvalue we will focus on two limiting cases.

\subsubsection{Case $v_a \to 0$} 

Let us first assume that the fermions belonging to the runaway/goldstino system 
do not mix with the other fermions, 
that is 
\be
v_a \to 0 \, . 
\ee
In this case the two fermions take part in a {\it seesaw} mechanism. 
Indeed we find the $c$-dependent eigenvalues of \eqref{MLL} to be 
\be
\label{PMGM}
\lambda_\pm = \frac12 \left( {\rm m} \pm \sqrt{ {\rm m}^2 + 4 (c/f)^2 } \right)  \, , 
\ee
along with the eigenvalues of ${\hat W}_{ab}$ that are independent of $c$. 
Clearly the effective goldstino mass will be related to $\lambda_\pm$. 
Let us see which of the two $\lambda_+$ or $\lambda_-$ is to be ascribed to the goldstino.

For $|$m$|\gg|c/f|$ we have 
\be
\label{mpm}
\lambda_+ \simeq {\rm m} + \frac{c^2}{{\rm m} f^2} \, , \quad \lambda_- \simeq - \frac{c^2}{{\rm m} f^2}  \, , 
\ee
up to order ${\cal O}(c^4)$. 
The effective goldstino mass is identified with the eigenvalue that vanishes when $c \to 0$, 
and is thus given by 
\be
\label{MGTT}
|m_G| \equiv |\lambda_-| =  \frac{c^2}{{\rm m} f^2} = \frac{V_\mathbb{B}}{M^2} \frac{1}{\text{m}} \, . 
\ee
We see that \eqref{MGTT} is different from \eqref{MGOO}. 
This happens because 
the goldstino is a linear combination of both fermions $\chi^o$ and $\chi^1$. 
The reader may wonder why the goldstino would receive contribution from the fermion $\chi^o$ when $f_o=0$. 
This happens because on the non-trivial background that we are studying we have 
\be
\dot A^o|_\mathbb{B} \ne 0 \, , 
\ee
and therefore the fermion $\chi^o$ will also shift under supersymmetry 
as can be seen from \eqref{SUSYGEN}. 
As a result, 
the scale m also enters into the eigenvalue $m_G$ characterizing the effective goldstino mass.

By studying the limit $c \to 0$ we thus found that it is the $\lambda_-$ 
eigenvalue of \eqref{PMGM} that relates to the goldstino. 
We can then also consider a different limit, namely $|$m$|\ll|c/f|$ and we see that  the effective goldstino mass is 
\be
\lambda_- = - \frac{c}{f} + \frac{{\rm m}}{2} + \frac{c}{f} \, {\cal O}({\rm m}^2 f^2 / c^2) \, . 
\ee
Interestingly, 
this limit reproduces the result of a single chiral superfield we have studied earlier. 
Indeed we find 
\be
\label{mfc}
|m_G| \simeq \frac{|c|}{|f|} =  \frac{\sqrt{V_\mathbb{B}}}{M} \, . 
\ee
The mass \eqref{mfc} seems not to vanish in the limit $c \to 0$. 
This happens because to derive \eqref{mfc} 
we have assumed that $|$m$|\ll|c/f|$, 
therefore we cannot apply the limit $c \to 0$ here unless m itself goes to zero, 
in which case $|m_G|$ would go to zero as well.

\subsubsection{Case $\frac{c}{f} \ll \mu$}

Another natural scenario that can allow us to study the mass matrix \eqref{M1} is to assume that 
\be
\label{assum}
\frac{c}{f} \ll \mu  \, , 
\ee 
where $\mu$ is a mass scale that characterizes collectively the {\it eigenvalues} of the matrices $\hat W_{ab}$ 
and $\begin{pmatrix}
  {\rm m} & v_a  \\ 
   v_b & {\hat W}_{ab}   
  \end{pmatrix}$. 
This is not an innocent assumption because there is no guarantee this will hold in realistic models, 
because of seesaw mechanisms for example, 
but it is a way that allows us here to proceed. 
In this limit we can recast the equation \eqref{MLL} to the form 
\be
\label{LGLG1}
\lambda_G = - \frac{c^2}{f^2} \frac{\det[\hat W_{ab} - \lambda_G \, \delta_{ab} ]}{\det\Big{[}\begin{pmatrix}
  {\rm m}-\lambda_G  & v_a  \\ 
   v_b & {\hat W}_{ab} - \lambda_G \, \delta_{ab}  
  \end{pmatrix}\Big{]}}  \, , 
\ee
and solve iteratively for $\lambda_G$ up to any order in $c/f$. 
This can be done by 
expanding the determinant in the numerator as 
\be
\label{dets}
\det[\hat W_{ab} - \lambda_G \, \delta_{ab} ] = \prod_i (\mu_i - \lambda_G) \, , 
\ee
where the $\mu_i$, with $\mu_i \sim \mu$, are the eigenvalues of the matrix $\hat W_{ab}$. 
A similar expansion can be performed for the determinant appearing in the denominator of \eqref{LGLG1}. 
Clearly the eigenvalue $\lambda_G$ in \eqref{LGLG1} describes the goldstino eigenvalue as it does go to zero when $c$ goes to zero. 
Let us stress that the iterative procedure is valid only because we are assuming the eigenvalues of the matrices appearing in 
\eqref{LGLG1} to be much larger than $c/f$, otherwise the procedure would not be a priori justifiable to give the correct $\lambda_G$. 
After the first step in the iterative procedure we see that the 
$\lambda_G$ eigenvalue will take the form 
\be {\ } \!\!\!\!\!\!\!\!\! 
\lambda_G = - \frac{c^2}{f^2} \frac{\det[{\hat W}_{ab}]}{\det \Big{[}\begin{pmatrix}
  {\rm m}  & v_a  \\ 
   v_b & {\hat W}_{ab}  
  \end{pmatrix}\Big{]}} 
  + \frac{c^2}{f^2 \mu} \, {\cal O}(c^2/f^2 \mu^2) \, .  
\ee 
Then the goldstino mass is of order  
\be 
\label{simp}
m_{G} \sim \frac{c^2}{f^2 \mu} \sim \frac{V_\mathbb{B}}{M^2 \mu} \, . 
\ee  
We see that here the goldstino mass behaves as in \eqref{MGTT}. 
If we also assume the $v_a$ to be much smaller than m and the eigenvalues of ${\hat W}_{ab}$, 
then the latter drop out and we find the $\lambda_-$ mass of \eqref{mpm}.

\subsection{${\hat W}_o =f_o \ne 0$ and ${\hat W}_A = f_A \ne 0$}

The possibility C. means that all the fermions would contribute to the goldstino. 
However, 
we can always rotate the superfields $\Phi^A$ such that only 
one of the rotated $\chi^A$ eventually contributes, 
therefore we have 
\be
f_o \ne 0  \, , \quad f_1 \ne 0 \, , \quad f_a = 0 \, . 
\ee
The conditions \eqref{V_0} and \eqref{V_A} imply 
\be
{\hat W}_{oo} f_o + {\hat W}_{1o} f_1 = c \, , \quad {\hat W}_{oA} f_o = - {\hat W}_{1A} f_1 \, . 
\ee
Once we set 
\be
{\hat W}_{oo} = {\rm m} \, , \quad {\hat W}_{oa} = v_a \, , 
\ee
the mass matrix takes the form 
\be
\label{M3}
m_{IJ} = 
\begin{pmatrix}
  {\rm m} &  \frac{c- {\rm m} f_o}{f_1}  & v_a  \\
 \frac{c- {\rm m} f_o}{f_1}   & \frac{{\rm m}f_o^2 - c f_o}{f_1^2} &- \frac{f_o}{f_1} v_a \\ 
   v_b & - \frac{f_o}{f_1} v_b & {\hat W}_{ab}  
  \end{pmatrix} \, . 
\ee
We search again for the mass eigenstate that will vanish in the limit $c \to 0$. 
We will study here only two limiting cases, 
and leave a detailed account of the properties of \eqref{M3} for a future work.

\subsubsection{Case $v_a \to 0$}

Let us first assume that the parameters that induce the mixing of the supersymmetry breaking sector 
with the matter fermions can be ignored, which means we set 
\be
v_a \to 0 \, . 
\ee
In this case the effective mass matrix eigenvalues for the supersymmetry breaking sector read 
\be
\lambda_\pm = \frac{{\cal B} }{2 } 
\pm  \frac{\sqrt{{\cal B}^2 - 4 \Big{[} \frac{c f_o}{f_1^2} {\rm m} -  \frac{c^2}{f_1^2} \Big{]} }}{2} \, , 
\ee
where 
\be
{\cal B} = \frac{f_o^2}{f_1^2} {\rm m} + {\rm m} - \frac{c f_o}{f_1^2} \, . 
\ee
The goldstino effective mass is given by the $\lambda_-$ because it goes to zero as $c$ goes to zero. 
For small $c/f_1$ we have 
\be
\lambda_- = \frac{c f_o}{f^2_1 + f_o^2} + {\cal{O}}(c^2) \, , 
\ee
which gives for the goldstino effective mass 
\be
\label{what}
m_G \simeq \frac{f_o}{M} \, . 
\ee
Again we see that the mass of the goldstino is proportional to the slope of the scalar potential.

\subsubsection{Case $\frac{c}{f} \ll \mu$}

A way to proceed without assuming that the $v_a$ are small is to shift the fermion $\chi^o$ as 
\be
\label{SFTSFT}
\chi^o \to \chi^o + \frac{f_o}{f_1} \chi^1 \, , 
\ee
which brings the mass matrix to the form 
\be
\label{M4}
m_{IJ} = 
\begin{pmatrix}
  {\rm m} &  \frac{c}{f_1}  & v_a  \\
 \frac{c}{f_1}   & \frac{c f_o}{f_1^2} & 0 \\ 
   v_b & 0 & {\hat W}_{ab}  
  \end{pmatrix} \, . 
\ee
Because of the shift \eqref{SFTSFT} the kinetic matrix of the fermions also changes and takes the form 
\be
\label{Kin}
k_{IJ} = 
\begin{pmatrix}
  1 &  \frac{f_o}{f_1}  & 0  \\
 \frac{f_o}{f_1}   &1+  \frac{f_o^2}{f_1^2} & 0 \\ 
   0 & 0 & \delta_{ab}  
  \end{pmatrix} \, . 
\ee
To find the effective mass eigenvalues we have to solve the eigenvalue equation $\det[m_{IJ} - \lambda \, k_{IJ} ] = 0$, 
that takes the form 
\be
\label{MLL'}
\begin{aligned}
&\left( \frac{c}{f_1} - \lambda \frac{f_o}{f_1} \right)^2 \, {\cal A}(\lambda) 
\\
&+ \left( \lambda + \lambda \frac{f^2_o}{f^2_1} - \frac{c f_o}{f^2_1} \right) \, {\cal B}(\lambda)  = 0 \, , 
\end{aligned}
\ee
where 
\be
\begin{aligned}
{\cal A}(\lambda) & = \det[\hat W_{ab} - \lambda \, \delta_{ab} ] \, , 
\\ 
{\cal B}(\lambda) & = \det\Big{[}\begin{pmatrix}
  {\rm m}-\lambda  & v_a  \\ 
   v_b & {\hat W}_{ab} - \lambda \, \delta_{ab}  
  \end{pmatrix}\Big{]} \, . 
\end{aligned}
\ee 

To proceed we will assume that \eqref{assum} holds also here for the scale $\mu$ that characterizes the 
eigenvalues $\mu_i$ of the matrices $\hat W_{ab}$ 
and $\begin{pmatrix} 
  {\rm m} & v_a  \\ 
   v_b & {\hat W}_{ab}   
  \end{pmatrix}$. 
In this limit we can solve \eqref{MLL'} iteratively 
by first bringing the determinants to the form \eqref{dets} and assuming $\mu_i \sim \mu$. 
Indeed we find  
\be
\!\!\!\!\!\!\!\! 
\lambda_G =  \frac{c f_o {\cal B}(\lambda_G) - c^2 {\cal A}(\lambda_G)}{ 
(f_1^2+f_o^2) {\cal B}(\lambda_G) 
+ ( \lambda_G f_o^2 - 2 f_o c ){\cal A}(\lambda_G) 
}  \, . 
\ee 
Clearly $\lambda_G$ reproduces the correct limit for $c \to 0$, 
and we can solve it up to any order in $c$. 
The eigenvalue $\lambda_G$ then reads 
\be
\lambda_G = 
\frac{c f_o }{ (f_1^2 + f_o^2 ) }  
  + {\cal O}(c^2) \, , 
\ee 
giving again \eqref{what} as the leading order contribution to the effective goldstino mass.

\section{Discussion}

Let us end this article with a discussion on the implications of our findings. 
We have demonstrated 
that in supersymmetric field theories with runaway potentials (or generically in time-dependent backgrounds) 
the goldstino instead of being massless 
becomes {\it massive}. 
This happens because the system is not stabilized {\it on} the vacuum, 
rather it is described by an effective theory around a non-trivial scalar background that evolves in time. 
As we have shown, 
the effective mass of such fermion 
has a specific order of magnitude under rather general assumptions. 
In the simplest examples it is controlled by the value of the scalar potential $V$ 
evaluated on the background and the scale $M$ that characterizes 
the slope 
\be
m_G \sim \frac{\sqrt{V}}{M} \, , 
\ee 
as we found in \eqref{mVM}. 
In a setup where the runaway fermion mixes with some other fermion via a Dirac mass, 
we would find a seesaw mechanism taking place, 
thus delivering 
\be
m_G \sim \frac{V}{M^2 \, \mu} \, ,
\ee
as for example we found in \eqref{MGTT}, 
where $\mu$ is a mass scale that enters the fermion mass matrix.

We did not study here gauged chiral models, 
but our results are expected to hold also when gaugings are introduced. 
The N=1 gauge multiplets do not have physical scalars, 
therefore the mass terms of the gaugini will be similar to those of the $\chi^A$ fermions 
appearing in \eqref{MASSterm}. 
This is however an interesting generalization that we leave for future work.

Let us also discuss if it is natural to have a small $c$ (we remind that $c=V'$). 
Recall that the pure Volkov--Akulov goldstino model enjoys a global R-symmetry, 
if we ignore higher order terms. 
However, 
the global R-symmetry is explicitly broken by a goldstino mass. 
Then, 
because the goldstino becomes strictly massless in the vanishing $c$ limit 
\be
m_G \Big{|}_{c \to 0} \to 0 \, , 
\ee 
the R-symmetry may in some cases be restored.  
Therefore, 
a small value for the goldstino mass would fall under the {\it technical naturalness} arguments \cite{tHooft:1979rat} 
as the symmetry of the system increases when $c \to 0$. 
As a result, 
for models where the R-symmetry is mildly broken by $c$, 
the massive goldstino will be 
protected from receiving a large mass from quantum corrections, 
and $c$ is naturally small.

An interesting question is how to proceed in the study of generic matter couplings 
and in particular how to construct low energy effective actions in the spirit 
of the non-linear realizations \cite{Clark:1996aw,Antoniadis:2010hs,Antoniadis:2012zz,Cribiori:2016hdz,Cribiori:2019cgz}. 
Such low energy effective theories 
can help to classify the generic 
couplings of the massive goldstino 
and point to possible 
distinct signals in experiments. 
In particular, 
such very light fermions 
could serve as dark matter that is weakly interacting 
with the matter particles of the Standard Model in the spirit of \cite{Liu:2016idz}. 
Extrapolating the properties of the massless goldstino, 
one would expect that 
the massive goldstino will couple to the Standard Model sector in a universal manner via the energy momentum 
tensor \cite{Clark:1996aw,Antoniadis:2010hs,Antoniadis:2012zz}. 
Namely we would have couplings of the form 
\be
\frac{i}{{\cal F}^4} (G \sigma^m \partial^n \overline G - \partial^n G \sigma^m \overline G ) \, T^{{\rm (SM)}}_{mn} \, , 
\ee
where $T^{(SM)}_{mn}$ is the energy momentum tensor of the Standard Model particles and 
${\cal F}$ some effective supersymmetry breaking scale. 
A detailed study of such fermion serving as a dark matter candidate is in order.

Finally, our findings are expected to 
hold also for extended globally supersymmetric theories. 
In particular, 
if the real scalar that appears in \eqref{LIN} transforms as 
\be
\delta \phi = \epsilon^i \psi_i + \dots 
\ee
then the arguments presented in the introduction would lead us to 
conclude that there exist non-trivial mass terms of the form 
\be
- c \, \psi_i G^i - c \, \overline \psi_i \overline G^i \, , 
\ee
where $G^i$ are the goldstini of the extended supersymmetric theory \cite{Cribiori:2016hdz}, 
transforming as $G^i \to G^i + \epsilon^i + \dots$ 
As a result massive goldstini would also arise in extended supersymmetry on runaway backgrounds. 
A detailed discussion of the fermion mass matrix of such theories is however beyond the scope of this work and we leave it for future research.

\section*{Acknowledgments} 

I would like to thank Nikolay Bobev, Niccol\`o Cribiori, Alex Kehagias and Thomas Van Riet for discussions. 
This work is supported from the KU Leuven C1 grant ZKD1118 C16/16/005.

\end{document}